\documentclass[aps,prc,twocolumn,showpacs]{revtex4}
\usepackage{graphicx,epsfig}
\def\PRL{Phys. Rev. Lett. }
\def\PRC{Phys. Rev. C }
\def\PRD{Phys. Rev. D }

\def\etal{\emph{et al.}}
\newcommand{ \be }{\begin{equation}}
\newcommand{ \ee }{\end{equation}}
\newcommand{ \bea }{\begin{eqnarray}}
\newcommand{ \eea }{\end{eqnarray}}
\newcommand{ \la }{\langle}
\newcommand{ \ra }{\rangle}

\newcommand{ \Dphi }{\Delta \phi}
\newcommand{ \ds }{\displaystyle}
\usepackage{color}

\begin{document}

\title{Elliptic flow contribution to two-particle correlations 
at different orientations to the reaction plane}
\author{J. Bielcikova$^1$,
S.~Esumi$^2$, K.~Filimonov$^3$, 
S.~Voloshin$^4$, and J.~P.~Wurm$^5$}

\affiliation{
$^1$ Physikalisches Institut, Heidelberg University, 69120 Heidelberg,
Germany\\ 
$^2$University of Tsukuba, Tsukuba, Ibaraki 305, Japan \\
$^3$Lawrence Berkeley National Laboratory, Berkeley, California
94720\\ 
$^4$Wayne State University, Detroit, Michigan 48201\\
$^5$Max-Planck Institut f\"ur Kernphysik, 69229 Heidelberg, Germany}

\date{\today}

\begin{abstract}
  
  Collective anisotropic particle flow, a general phenomenon present 
  in relativistic heavy-ion collisions, can be separated from direct particle-particle
  correlations of different physics origin by virtue of its specific
  azimuthal pattern.  We provide expressions for flow-induced
  two-particle azimuthal correlations, if one of the particles is
  detected under fixed directions with respect to the reaction plane.
  We consider an ideal case when the reaction plane angle is exactly
  known, as well as present the general expressions in case of finite
  event-plane resolution.  We foresee applications for the study of
  generic two-particle correlations at large transverse momentum
  originating from jet fragmentation.
\end{abstract}


\maketitle

\section{Introduction}

Collective particle flow is a general phenomenon of relativistic
heavy-ion collisions that originates from pressure gradients built up
in the anisotropic overlap zone of colliding
nuclei~\cite{ollitrault92}. Azimuthal anisotropies in inclusive {\it
single} particle distributions relative to the reaction plane
(anisotropic flow) have been extensively
studied~\cite{e877,na49,ceres,star,phenix,phobos}.  Recent
investigations of direct {\it two (or more)} particle correlations also
indicate that the dependence of these correlations on the orientation of
the reaction plane may contain important physics information.
A detailed analysis of such correlations requires flow effects
to be taken into account.

A recent example, which gave the motivation for this paper, is
provided by azimuthal two-particle correlations at transverse momenta
above 1~GeV/c. Such particles presumably originate from fragmentation
of dijets, but are embedded in collective flow~\cite{ceres}. It is
predicted that nuclear effects may modify the jet fragmentation
function due to induced radiation of the leading
parton~\cite{quenching}. This could result in significant changes in
the particle correlations within the jet, as well as the correlation
of particles originating from back-to-back jets. The modifications 
of the jet profile may depend on the nuclear geometry
and could be studied relative to the reaction plane
angle~\cite{ceres}.

In this paper, we present analytical formulae for 
the flow contribution to two-particle
azimuthal distributions for different orientations of the trigger
particle with respect to the reaction plane, neglecting any non-flow
effects. 
We will first discuss an ideal case with the reaction plane angle 
exactly known and then
incorporate the finite resolution of the reconstructed event plane.

\section{Anisotropic transverse flow}
Anisotropic flow manifests itself by the presence of higher ($n \ge 1$) 
harmonics  in the inclusive single particle
distribution in the azimuthal angle $\phi$ with respect to the reaction
plane $\Psi_R$~\cite{e877,vol}:
\be
\frac{dN}{d(\phi-\Psi_R)} \propto 
(1+\sum_{n=1}^\infty 2\; v_n \cos (n(\phi-\Psi_R))).
\label{single-rp}
\ee
The Fourier coefficients, $v_n=\la \cos(n(\phi-\Psi_R))\ra$, given by
the average over detected particles in analyzed events
quantify the anisotropy of the $n-$th harmonic of the distribution.
The anisotropies corresponding to the first and the second 
Fourier coefficients, $v_1$ and $v_2$, are usually
referred to as directed and elliptic flow, respectively.

Collective flow generates azimuthal anisotropies also in
the angle difference $\Delta \phi =\phi_i-\phi_j$ ($0\leq\Delta\phi\leq\pi$)
of particle {\it pairs}~\cite{pair},
\begin{equation}
\!\frac{dN^{pairs}}{\pi \, d\Delta\phi}\!=B\,(1\!+\!\sum_{n=1}^\infty 2 \;p_n (p_{Ti},y_i; p_{Tj},y_j) 
\cos(n \Delta \phi)), 
\label{pair}
\end{equation}
where $B$ denotes the integrated inclusive pair yield.
In case of pure collective flow, the Fourier coefficients $p_n=\la
\cos(n\Delta\phi)\ra$ are given by~\cite{posk}
\be
p_n(p_{Ti},y_i; p_{Tj},y_j) = v_{n}(p_{Ti},y_i)\, v_{n}(p_{Tj},y_j).
\label{pall}
\ee

\section{Pair distributions in $\Dphi$ when 
the trigger particle is detected at fixed angle relative to the event plane}

\begin{figure}[t!]
\includegraphics[width=0.45\textwidth]{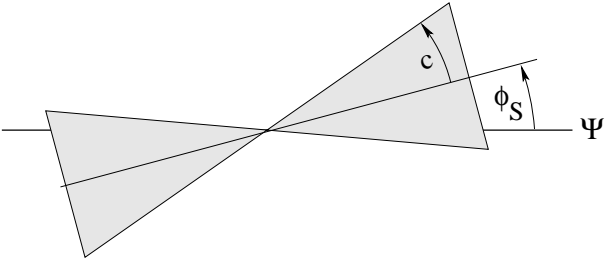}
\caption{\label{conus}
  The region ${\cal R}$ is made up of a bi-sector of half-angle $c$
  that intersects the reaction plane $\Psi$ at angle $\phi_S$, {\it
  modulo} $\pi$.}  
\end{figure}

We introduce conditional two-particle correlations in the transverse
plane for which one of the particles, usually referred to as the {\it
trigger particle}, is detected within some {\it bi-sector} ${\cal R}$ at
fixed orientation with respect to the reaction plane, see
Fig.~\ref{conus}. 

The $n$-th harmonic of the pair distribution, before given by 
Eq.~(\ref{pall}), is expressed as
\be
p_n^{\cal R}=v_{n}(p_T,y)\;v_{n}^{\cal R}(p_T,y).
\label{pnR}
\ee
To simplify the notations, we have assumed that both particles are
detected in the same $p_T$ and $y$ interval, but it is straightforward
to generalize our results for the case when the trigger particle and
the associated particle are chosen from different rapidity and
transverse momentum regions.  Here, $v_{n}^{\cal R}= \la
\cos(n(\phi-\Psi)) \ra^{\cal R}$ is the $n$-th harmonic coefficient of
the single-particle distribution of Eq.~(\ref{single-rp}), although the
average over the azimuthal angle of the trigger particle is 
taken over the restricted region ${\cal R}$ only.

We derive now explicit analytic
expressions for $v_{2}^{\cal R}$ and the pair yield $B^{\cal R}$ for
elliptic flow when the trigger particle is confined to a {\it bi-sector}
oriented with angle $\phi_S$ to the reaction plane, and then
specialize to {\it in-plane} and {\it out-of-plane} conditions. We
proceed in two steps, first for the ideal case, then for finite resolution
in the reconstructed event plane.

\subsection{Ideal case. Reaction plane is known. }

Let the trigger particle be confined in the transverse plane to the
{\it bi-sectors} depicted in Fig.~\ref{conus}. The $n$-th Fourier
coefficient of the trigger particle distribution, assuming it is
originally given by Eq.~(\ref{single-rp}), is 
\begin{widetext}
\begin{eqnarray}
v_{n}^{\cal{R}}
&=&
\la \cos (n(\phi-\Psi_R))\ra^{\cal{R}}
=
\frac{\int\limits_{\cal{R}}(1+\sum\limits_{k=1}^\infty
2v_k\cos(k(\phi-\Psi_R))) 
\cos(n(\phi-\Psi_R))d(\phi-\Psi_R)}
{\int\limits_{\cal{R}}(1+\sum\limits_{k=1}^\infty 
2v_{k}\cos({k}(\phi-\Psi_R)))d(\phi-\Psi_R)}, 
\label{int-vnr-rp}
\end{eqnarray}
where the integration over the region ${\cal R}$ in more explicit
notation is understood to read
\be
\int\limits_{\cal{R}} d(\phi-\Psi_R)\; .\;.\;. 
\equiv \int\limits_{\phi_S-c}^{\phi_S+c} d(\phi-\Psi_R)\;.\;.\;. + 
\int\limits_{\phi_S+\pi-c}^{\phi_S+\pi+c} d(\phi-\Psi_R)\; .\;.\;. \;\;.
\label{schematic-integration}
\ee
The integration results in  
\begin{equation}
v_n^{\cal R}=
\frac{v_n+\delta_{n,even}\cos(n\phi_S)\frac{{\textstyle\sin(nc)}}
{\textstyle{nc}}+
\sum\limits_{k=2,4,6,...}({\textstyle v_{k+n}+v_{|k-n|}})\cos(k\phi_S)
\frac{\textstyle {\sin(kc)}}{\textstyle{kc}}}
{1+\sum\limits_{k=2,4,6,...}     
{\textstyle 2\,v_k}\cos(k\phi_S)\frac{\textstyle{\sin(kc)}}
{\textstyle{kc}}},
\label{vnr-rp}
\end{equation}
where $\delta_{n,even}$~=~1 for $n$ even and $\delta_{n,even}$~=~0 
for $n$ odd, respectively.
\end{widetext}

Spatial conditions on the trigger particle also modify the integrated
pair yield. We express the conditional two-particle yield
as 
\begin{equation}
B^{\cal R}= \frac{2c}{\pi}\,B\,\beta^{\cal R},
\label{B-beta}
\end{equation}
which can be understood as the product of two single-particle yields:
$\sqrt{B}$ for the associated particle and the remainder
$\sqrt{B}\,\frac{\textstyle{2c}}{\textstyle{\pi}}\,\beta^{\cal R}$
for the trigger particle. Here, $\frac{\textstyle{2c}}{\textstyle{\pi}}$
is the fraction of the azimuth covered by the trigger particle and
the quantity $\beta^{\cal R}$ accounts
for the modification of the yield due to collective flow and
is given by
\begin{equation}
\beta^{\cal R} = \frac{ {\int\limits_{\cal{R}}(1+
    \sum\limits_{k=2,4,6,...} 
    2v_k\cos (k(\phi-\Psi_R)))d(\phi-\Psi_R)}}
    {\int\limits_{\cal{R}}\,d(\phi-\Psi_R ) }.
\label{int-betar-rp}
\end{equation}
Integrating we obtain
\begin{equation}
\beta^{\cal R}=
1+\sum\limits_{k=2,4,6,...} 2\,v_k\cos(k\phi_S)\frac{\textstyle \sin(kc)}{kc}.
\label{betar-rp}
\end{equation}
In the following we restrict ourselves to elliptic flow ($n$~=~2).
Neglecting terms with $n\ge$~4, we obtain 
\begin{equation}
v_{2}^{\cal{R}}=\frac{v_2+\ds\cos(2\phi_S)\frac{\textstyle{\sin(2c)}}{\textstyle{2c}}
+v_2\cos(4\phi_S)\frac{\textstyle{\sin(4c)}}{\textstyle{4c}}}
{1+\ds 2 v_2\cos(2\phi_S)\frac{\textstyle{\sin(2c)}}{\textstyle{2c}}},
\label{v2r-rp}
\end{equation}
and
\begin{equation}
\beta^{\cal{R}}= 1+ 2v_2\cos(2\phi_S)\,\frac{\textstyle{\sin(2c)}}{\textstyle{2c}}.
\label{beta2r-rp}
\end{equation}

If the trigger particle is confined to regions $-\pi/4 < \phi-\Psi_R <
\pi/4 $ ($\phi_S$~=~0, {\it 'in-plane'}~), and $\pi/4 < \phi-\Psi_R < 3\pi/4$ 
($\phi_S= \pi/2$,~{\it 'out-of-plane'}~), respectively, Eq.~(\ref{v2r-rp}) 
simplifies to
\begin{equation}
v_2^{\rm in}=\frac{\pi v_2+2}{\pi+4v_2},~~~  v_2^{\rm out}
=\frac{\pi v_2-2}{\pi-4v_2}.
\label{vinout-rp}
\end{equation}

\begin{figure}[t!]
\includegraphics[width=8.0cm]{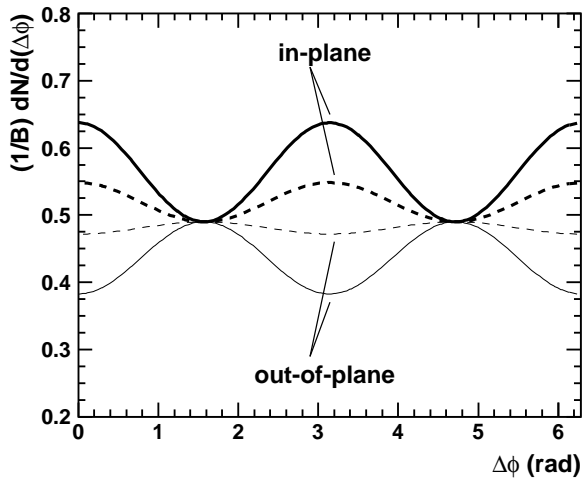}
   \caption{{\it In-plane} and {\it out-of-plane}
   correlation functions for ideal reaction plane (full lines), 
   and for finite event plane resolution 
   ($\la\cos(2\Delta\Psi)\ra=0.3$) (dashed lines).
   The trigger particle 
   is confined to bi-sectors with
   axes $\phi_S$ pointing along the reaction plane ($\phi_S= \Psi$)
   and perpendicular to it ($\phi_S= \Psi+\pi/2$), respectively.
   The magnitude of elliptic flow is $v_2$~=~10\%.      
   } 
\vspace{0.5cm}
\label{distributions}
\end{figure}

The pair yields under these conditions are 
\begin{equation}
B^{\rm in}=\frac{B}{2}\left(1+\frac{4v_2}{\pi}\right),~~~
B^{\rm out}=\frac{B}{2}\left(1-\frac{4v_2}{\pi}\right)
\label{binout-rp}
\end{equation}
which add up to $B$ as both regions cover together the full azimuth.

The azimuthal distributions for in and out-of-plane conditions are
obtained by inserting the corresponding expressions for
$v_{2}^{\cal{R}}$ into Eq.~(\ref{pnR}) and then $p_2^{\cal{R}}$ and
$B^{\cal{R}}$ into Eq.~(\ref{pair}). The normalized in-plane and
out-of-plane distributions for $v_2$~=~0.1 are displayed in
Fig.~\ref{distributions} (full line). The out-of-plane distribution is
shifted in phase by $\pi/2$ compared to the in-plane distribution:
instead of peaks at $\Delta\phi=$~0 and $\pi$ peaks show up at
$\pi/2$ and $3\pi/2$. The sign of $v_2^{\rm out}$ is negative.  Both
curves touch at level $(B/2)(1-2v_2^2)$.

\subsection{Finite event plane resolution}

The direction of the true reaction plane $\Psi_R$ is not available
experimentally.  An estimator for the reaction plane, often called the
event plane, $\Psi_E$, is determined event-by-event using the
anisotropic flow itself~\cite{posk}. How close on average the event
plane is to the true reaction plane is determined by the resolution,
usually quantified by $\la\cos(n\Delta\Psi)\ra$, where
$\Delta\Psi=\Psi_E-\Psi_R$.  Here, the angular brackets 
$\la\cdots\ra$ indicate the event averaging over the probability density 
distribution $\rho(\Delta\Psi)$ that characterizes 
the event plane resolution.

\begin{figure}[b!]
\includegraphics{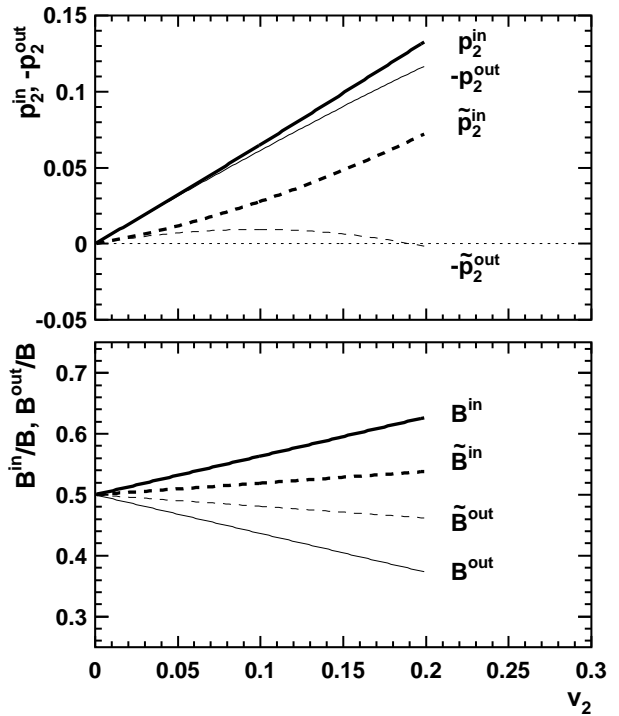}
\caption{In-plane (thick lines) and out-of-plane coefficients (thin lines) 
$p_2$ of Eq.~(\ref{pnR}), (top), and $B$ of Eq.~(\ref{B-beta}),
(bottom), {\it vs} elliptic flow anisotropy $v_2$. Solid lines assume
ideal reaction plane, dashed lines are for reconstructed event planes
with finite resolution $\la\cos(2\Delta\Psi)\ra=0.3$.}
\label{inout-v2-rp-ep}
\end{figure}
Let us now calculate how the finite event plane resolution modifies
our results. For a given deviation $\Delta\Psi$ the new range of
integration $\widetilde{\cal{R}}$ in Eq.~(\ref{int-vnr-rp}) and
Eq.~(\ref{int-betar-rp}) is defined in analogy to
Eq.~(\ref{schematic-integration}) by
\begin{widetext}
\be
\int\limits_{\widetilde{\cal{R}}} d(\phi-\Psi_R)\;
 .\;.\;. \equiv \int\limits_{\phi_S+\Delta\Psi-c}^{\phi_S+\Delta\Psi+c} 
d(\phi-\Psi_R)\;.\;.\;. + \int\limits_{\phi_S+\Delta\Psi+\pi-c}^{\phi_S+\Delta\Psi+
\pi+c} d(\phi-\Psi_R)\; .\;.\;. \;\; .
\ee
The $n$-th Fourier harmonic component is obtained after averaging over the
probability density distribution $\rho(\Delta\Psi)$,
\begin{equation}
\widetilde{v_n^{\cal{R}}}=
\frac{\int\limits_{-\pi}^{\pi} 
\rho(\Delta\Psi)
\;\int\limits_{\widetilde{\cal{R}}}(1+\sum\limits_{k=1}^\infty 2v_k
\cos (k(\phi-\Psi_R)))
\cos(n(\phi-\Psi_R))d(\phi-\Psi_R)\;d(\Delta\Psi)}{\int\limits_{-\pi}^{\pi} 
\rho(\Delta\Psi)
\;\int\limits_{\widetilde{\cal{R}}}
(1+\sum\limits_{k=1}^\infty 2v_k\cos(k(\phi-\Psi_R)))d(\phi-\Psi_R)\;d(\Delta\Psi)}.
\label{int-vnr-ep}
\end{equation}
After integration we obtain
\be
\widetilde{v_n^{\cal R}}=
\frac{v_n+\delta_{n,even}\cos(n\phi_S)\frac{{\textstyle\sin(nc)}}{\textstyle{nc}}\la\cos(n\Delta\Psi)\ra+
\sum\limits_{k=2,4,6,...}({\textstyle v_{k+n}+v_{|k-n|}})\cos(k\phi_S)\frac{\textstyle {\sin(kc)}}{\textstyle{kc}}\la\cos(k\Delta\Psi)\ra}
{1+\sum\limits_{k=2,4,6,...}     
{\textstyle 2\,v_k}\cos(k\phi_S)\frac{\textstyle{\sin(kc)}}{\textstyle{kc}}\la\cos(k\Delta\Psi)\ra}.
\label{vnr-ep}
\ee
In analogy, we can write
\begin{equation}
\widetilde{\beta^{\cal R}} = \frac{\int\limits_{-\pi}^{\pi}
\rho(\Delta\Psi)
\int\limits_{\widetilde{\cal{R}}}
(1+\sum\limits_{k=2,4,6,...} 2 v_k\cos(k(\phi-\Psi_R)))d(\phi-\Psi_R)d(\Delta\Psi) }
    {\int\limits_{\widetilde{\cal{R}}}   \,d(\phi-\Psi_R ) }.
\label{int-betar-ep}
\end{equation}
After integration we obtain:
\begin{equation}
\widetilde{\beta^{\cal R}}=1+\sum\limits_{k=2,4,6,...}   
\textstyle 2\,v_k\cos(k\phi_S)\frac{\textstyle{\sin(kc)}}{\textstyle{kc}}\la\cos(k\Delta\Psi)\ra.
\label{betar-ep}
\end{equation}

In the following we restrict ourselves again to elliptic flow ($n=$~2)
only, and neglecting terms with $n~\ge$~4, we obtain
\begin{eqnarray}
\widetilde{v_2^{\cal{R}}}
&=&
\frac{v_2+\ds\cos(2\phi_S)\frac{\textstyle{\sin(2c)}}{\textstyle{2c}}\la\cos(2\Delta\Psi)\ra
+v_2\cos(4\phi_S)\frac{\textstyle{\sin(4c)}}{\textstyle{4c}}\la\cos(4\Delta\Psi)\ra}
{1+\ds 2v_2\cos(2\phi_S)\frac{\textstyle{\sin(2c)}}{\textstyle{2c}}\la\cos(2\Delta\Psi)\ra},
\label{v2r-ep}
\end{eqnarray}
\end{widetext}
and
\begin{equation}
\widetilde{\beta^{\cal{R}}}= 1+2v_2\cos(2\phi_S)\,\frac{\textstyle{\sin(2c)}}{\textstyle{2c}}\la\cos(2\Delta\Psi)\ra.
\label{beta2r-ep}
\end{equation}

The {\it in-plane} and {\it out-of-plane} anisotropies of
Eq.~(\ref{vinout-rp}) for elliptic flow
are modified for finite event plane resolution to
\begin{eqnarray}
\widetilde{v}_2^{\rm in} &=& \frac{\pi v_2+2\la\cos(2\Delta\Psi)\ra}
{\pi+4v_2\la\cos(2\Delta\Psi)\ra}, \nonumber \\
\widetilde{v}_2^{\rm out}&=& \frac{\pi v_2-2\la\cos(2\Delta\Psi)\ra}
{\pi-4v_2\la\cos(2\Delta\Psi)\ra},
\label{vinout-ep}
\end{eqnarray}
and the average yields of Eq.~(\ref{binout-rp}) to
\begin{eqnarray}
\widetilde{B}^{\rm in} &=& \frac{B}{2}\left[1+\frac{4v_2}{\pi}\la\cos(2\Delta\Psi)\ra\right],
\nonumber \\
\widetilde{B}^{\rm out}&=& \frac{B}{2}\left[1-\frac{4v_2}{\pi}\la\cos(2\Delta\Psi)\ra\right],
\label{binout-ep}
\end{eqnarray}
respectively. These formulae have been used to calculate the dashed lines
in Fig.~\ref{distributions}, and it is seen
that the magnitude of the elliptic anisotropy is reduced for 
finite event plane resolution. The normalized background
parameters $\widetilde{B}^{\rm in}/B$ and $\widetilde{B}^{\rm out}/B$ 
approach the value of 0.5. Both are consequences
of the finite event plane resolution which causes the
in-plane region to receive also negative contributions from the
out-of-plane region, and vice versa.

Fig.~\ref{inout-v2-rp-ep} presents a synopsis of the dependence of the
flow parameters under {\it in-plane} and {\it out-of-plane} conditions
on the magnitude $v_2$ of elliptic flow, both for ideal as well as for
the reconstructed event plane. For the latter case, the reaction plane
resolution was chosen to be $\la\cos(2\Delta\Psi)\ra=0.3$.  Note that
very large $v_2$ and small $\la\cos(2\Delta\Psi)\ra$ could lead to the
situation of $v_2^{\text{out}}>0$, and the phases of {\em in-plane}
and {\em out-of-plane} distributions, Fig.~\ref{distributions}, would
be the same.

\section{SUMMARY AND OUTLOOK}
We have presented general expressions of two particle azimuthal
correlations due to anisotropic flow for the case when one of the
particles, referred to as the trigger particle, is detected at fixed
angles relative to the reaction plane. Analytical formulae are given
for two cases, an ideal case when the reaction plane is exactly known
in every event, and for the case of finite reaction plane
resolution. For the so called in-plane and out-of-plane conditions, we
find that the correlation functions are shifted in phase by $\pi/2$
for realistic values of elliptic flow of the trigger particle and the
reaction plane resolution. This and the increase in modulation
amplitude {\it in-plane} compared to {\it out-of-plane} is easily
visualized by the fact that the trigger particle scans the peak region
of the elliptic flow pattern in the first case, but the valley in the
second.

We foresee that the results presented in this paper will allow to
disentangle non-flow generic two-particle correlations, like those due
to jets and analyze how such correlations depend on the orientation of
the jet with respect to the reaction plane.

\acknowledgments{We are grateful to Ulrich Heinz for a critical reading
of the manuscript and clarifying suggestions. One of us (J.B.) gratefully 
acknowledges continuous interest and support by Johanna Stachel.
This work was supported in part by the U.S. Department of Energy under
Contract DE-AC03-76SF00098 and DE-FG02-92ER40713.
\bibliographystyle{unsrt}

 \end{document}